\documentstyle[12pt,epsf]{article}
\newcommand{\heading}[1]{\vspace*{15mm}
{\Large\begin{center} {\bf{#1}} \end{center}}}
\renewcommand{\author}[3]{\vspace{5mm}  
\begin{center}
{\normalsize \rm #1}\\    
{\normalsize \it #2}\\    
\vspace{0.65cm}
\framebox[3.8truecm]{\rule[-1.9cm]{0.cm}{3.8truecm}}
\vspace{0.35cm}
\end{center} }
\newcommand{\acknowledgements}[1]{\vspace{7mm} \noindent 
        {\normalsize \bf Acknowledgements.\,} {\normalsize #1}}
\def \aa#1#2   {{\em Astr. Astrophys. \/} {\bf #1}, {#2}}
\def \aas#1#2  {{\em Astr. Astrophys. Suppl. Ser. \/} {\bf #1}, {#2}}
\def \aj#1#2   {{\em Astron. J. \/} {\bf #1}, {#2}}
\def \apj#1#2  {{\em Astrophys. J. \/} {\bf #1}, {#2}}
\def \apjs#1#2 {{\em Astrophys. J. Suppl. Ser. \/} {\bf #1}, {#2}}

\def \nat#1#2  {{\em Nature \/} {\bf #1}, {#2}}
\begin{document}
\heading{COMPARISON OF {\it COBE} DMR AND\\ {\it ROSAT} ALL-SKY SURVEY 
DATA\footnote{to appear, Proceedings of the XXXIth Moriond Meeting
``Microwave Background Anisotropies'',\\ Les Arcs (France), 
March 16-23 1996}} 

\author{R\"udiger Kneissl} {Max-Planck-Institut f\"ur Astrophysik, Garching
bei M\"unchen, Germany}

\begin{abstract}{\baselineskip 0.4cm 
Statistical comparisons of microwave maps in the GHz range and
X-ray maps at around 1 keV are an interesting probe to constrain 
different astrophysical phenomena. Possible correlations on various 
angular scales and with different frequency (energy) dependences, 
although not expected at present day experimental sensitivity, could
in principle be due to Galactic emission/absorption, the Sunyaev-Zel'dovich 
effect, the Integrated Sachs-Wolfe effect in cosmological models 
with a cosmological constant or low density, or X-ray luminous radio
sources such as radio-loud AGNs. I report on work cross-correlating 
the COBE DMR and ROSAT All-Sky Survey in a selected area of the sky. 
This area (\mbox{$+\,40^\circ <$ b}, \mbox{$70^\circ <$ l $< 250^\circ$}) 
is the best presently available data set probing the medium-hard extragalactic 
X-ray background around 1 keV. 
No significant correlation on astrophysically relevant 
scales has been found in this analysis, but it will be possible to 
infer constraints from the limits.} 
\end{abstract}

\newpage
\section{Introduction}
Statistical comparison between experiments probing different astrophysical 
backgrounds are motivated by physical effects leading to possible
correlations. These effects, galactic emission or absorption in
X-rays, Sunyaev-Zel'dovich effect (SZ) in hot plasmas such as galaxy 
clusters or group halos, point sources, and the
Integrated Sachs-Wolfe effect (ISW), are astrophysically
interesting (see also Table 1) and test sources of possible confusion 
for the measurement of the cosmic microwave (CMB) and X-ray (XRB)
backgrounds.\\
Comparisons between CMB and XRB observations were first carried out by 
Boughn \& Jahoda \cite{boujah93} comparing the 19 GHz
map with HEAO-1 A2 ($\sim$10 keV, $3^\circ$ resolution), and they found no 
significant correlation based on Monte Carlo simulations for noise 
properties. Bennett et al. \cite{benetal93} cross-correlating 
the 1 yr DMR data to HEAO-1 found no significant correlation 
for $|b| > 30^\circ$ and the LMC 
masked. In the latest analysis by Banday et al. \cite{banetal96} an 
expansion in orthogonal functions on a cut sky and a likelihood analysis 
for the coupling constant between the 4 year DMR and HEAO-1 data was
used. Again no significant correlation was found, when applying a specially 
designed Galactic cut based on correlations obtained from 
the Dirbe 140 $\mu$m map and masking the LMC. \\
The present analysis uses another X-ray data set, the ROSAT
\cite{tru83} PSPC \cite{pfeetal86} All-sky Survey (RASS) which 
measures the sky in the soft X-rays from 0.1 -- 2 
keV with an angular resolution of $\sim$10 arcmin. 
These data have not previously been used for this kind of analysis and are 
particularly sensitive to galaxy clusters even in the diffuse XRB 
component, as has been shown by Soltan et al. \cite{soletal95}. 
They find an extended X-ray component around clusters. The present 
work is also motivated towards constraining a possible extended gas halo.\\
\section{The COBE and ROSAT data}
A full description of the COBE DMR data and the 4 year results can be
found in the contribution by G.F. Smoot \cite{smothese} and references
therein.\\
For our analysis we used various maps of the COBE DMR data. The final 
analysis uses the 4 year data, although consistency tests with the 1 year and 
2 year data were performed. From the different channels we
constructed inverse noise weighted A+B sum maps.
The same procedure was used for combining the different frequency
channels (31.5, 53 \& 90 GHz), which were converted from antenna to Planck
temperatures. The standard frequency combination for 
this analysis is the 53+90 GHz map, having low noise and little galactic 
contribution \cite{kogetal96}, but the individual frequency maps and 
the linearly combined and subtracted galaxy reduced maps were used for 
comparison. \\
The RASS R6 energy band, 0.73 -- 1.56\,keV with maximum
response around 1.1\,keV, is regarded as the best probe for the 
diffuse cosmological XRB because the content of foreground, 
non-cosmic photons, and contamination by 
charged particles is minimized. In spite of the careful corrections 
for exposure and elimination of non-cosmic backgrounds
\cite{snoetal95} the final count rate distribution 
is not completely free from residual contamination.
To test for spectral dependence we compared to the neighbouring 
partly overlapping band R5 (maximum response around 0.9\,keV), 
which is low in non-cosmic photons and in contamination by charged
particles, but contains increased galactic foreground.
The maps supplied for this work were binned as 
0.7$^\circ$ x 0.7$^\circ$, with point sources left in to compare to 
the complete integrated flux. The mean 
intensity of the XRB in the R6 is $\sim 80 \times 10^{-6}$
cts s$^{-1}$ arcmin$^{-2}$ with a fluctuation level of $\sim 7
\times 10^{-6}$ cts s$^{-1}$ arcmin$^{-2}$ at the COBE pixelization level 6 
(2$^\circ.6$ x 2$^\circ.6$).
Even the high energy R6 band is, in large regions of the sky, 
dominated by galactic emission. The chosen area 
(\mbox{$+\,40^\circ <$ b}, \mbox{$70^\circ <$ l $< 250^\circ$}) 
is the largest simply connected patch probing dominantly the XRB. 
Properties of the XRB in this field have been studied in a series of
papers (\cite{soletal95},\cite{miyetal96},\cite{soletal96}). 

\section{Correlation Method}
Due to the small size ($\sim 8\%$ of the sky) and the peculiar 
geometry of the patch a local statistical measure, 
the 2-point cross-correlation function, is preferred. 
The form of the correlation function used in this analysis is the 
Pearson product moment correlation coefficient 
\begin{equation} 
C(\alpha) = \frac{<X_i T_j>_\alpha \, - \, <X_i>_\alpha 
\, <T_j>_\alpha}{\sqrt{<X_i^2>_\alpha \, - \, <X_i>_\alpha^2} \,\,\,  
\sqrt{<T_j^2>_\alpha \, - \, <T_j>_\alpha^2}} 
\end{equation}
in an unweighted scheme. Inverse noise variance weighting has also been 
applied and was found to give similar results. The average is taken over 
all pixel pairs $\{ij\}$ with separation $\alpha$ in the patch. The 
subscript $\alpha$ denotes that all the terms are evaluated
separately for each angular bin. 
To determine the uncertainties, which are assumed to be dominated by
the DMR noise on small scales and the cosmic variance of the CMB
structure on large scales, different techniques were applied. These are
a simple rotation method using random samples, and simulation of DMR maps.
The CMB structure is taken to be a random Gaussian field with a power law 
($Q_{rms-PS}$ = 15.3 $\mu$K, n = 1.2) power spectrum \cite{goretal96} 
convolved with the DMR filter function \cite{knesmo93} and the 
DMR noise to be Gaussian pixel noise distributed according 
to the coverage.\\

\section{Results}
\begin{figure}[htbp]
\epsfxsize=0.9\hsize
\epsffile{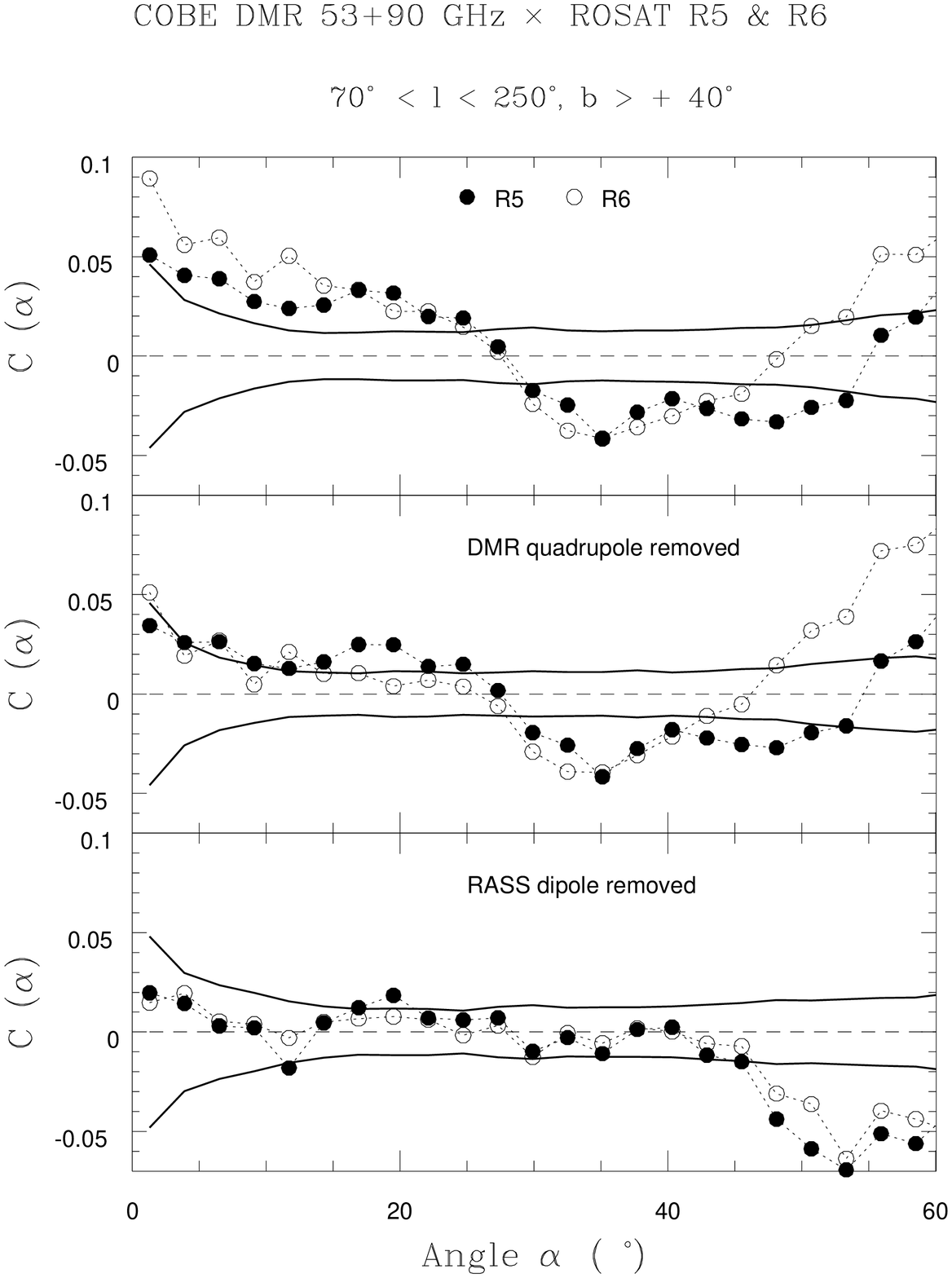}
\caption{Cross-correlation function between the COBE DMR 53+90 GHz
(A+B)/2 and the ROSAT All-sky survey energy bands R5 (softer band, open 
circles) and R6 (harder band, full circles) in a ROSAT selected area 
(\mbox{$+\,40^\circ <$ b}, \mbox{$70^\circ <$ l $< 250^\circ$}).
The effect of 
lowering the zero-lag amplitude by subtracting either a best fit quadrupole 
from the Galactic cut DMR map or a best fit dipole which has Galactic 
signature from the full ROSAT map can be seen. The 1-$\sigma$ error
bands are taken from DMR simulations correlated to R6.} 
\end{figure}
Correlating the raw data yields a marginally significant positive 
correlation on large angular scales, which appeared to be independent 
of different procedures that had been applied. 
Different source exclusion thresholds in the ROSAT maps ranging 
from 0.3 - 1 cts s$^{-1}$  were compared. 
By far the strongest source in the field with 5.3 cts s$^{-1}$ in the
R6 is MKN 421, a BL Lac object at 
z $\sim$ 0.031, which had to be removed, because it
clearly produces positive correlation on the beam scale. 
Otherwise the different thresholds affect the results only marginally. 
Different sampling tests were undertaken also showing
stability of the result against small scale features such as point
sources and noise. The maps were smoothed on various angular scales
including smoothing of the ROSAT maps with the actual DMR beam
\cite{knesmo93}, and Gaussian
smoothing of both maps out to $20^\circ$ with the effect of smoothing 
the correlation function, but not significantly changing the
correlated signal. 
Correlating to noise maps revealed no correlated noise feature.\\
The energy dependence in X-rays is found to be consistent with a 
galactic signal increasing from hard to soft energies. 
The frequency dependence in microwaves is unclear. 
There is a clear signal in both the 
53 and 90 GHz channels and no signal at 31.5 GHz.\\
To determine the angular scale of the correlated signal, gradients were 
removed from the field. This was done in fitting dipoles onto 
the field in both maps and subtracting them. As a result the signal 
vanished. The multipoles on the sky dominating these gradients 
turn out to be of low order (Figure 1). The best fit DMR residual
dipole, which would introduce substantial correlation of no physical 
significance had been removed in addition to the standard removed dipole. 
The gradient in the ROSAT field is 
produced by a whole map dipole which has the following galactic
signatures, the positive pole lying near 
the galactic center, and increasing relative amplitude from the hard to 
the soft energies, also in comparison to R7 and R4. The gradient in the COBE 
field is dominated by a quadrupole fitted to the COBE cut sky, a
combination of the cosmic and the Galactic quadrupole. 
In the field the cosmic quadrupole seems to dominate, not inconsistent 
with the COBE frequency dependence of the correlated signal. \\

\section{Discussion}
The conclusion of this work is that no significant correlation 
on scales from 7$^\circ$ to $60^\circ$ has been found between the 
COBE DMR and the ROSAT All-Sky Survey in the area investigated. A possible 
correlation on larger scales can not be distinguished from chance alignment, 
because the corresponding multipole terms are different and the frequency and 
energy dependences are not conclusive for a single physical mechanism. 
The full description of this work, also including the 
analysis of larger areas of the sky, as well as results on a ROSAT dipole 
determination and limits on the correlation coefficient with a
discussion of the possible mechanisms, can be found in a paper by 
our collaboration \cite{kneetal96}.\\
Different effects resulting in a possible, although not at present expected, 
angular correlation between microwave and X-ray experiments are known 
or have been suggested recently.\\

\small
\begin{table}[htb]
\begin{tabular}{|l|c|c|c|c|l|}
\hline
 {\it EFFECT /} & {\it SIGN} & {\it ANGULAR} & {\it
FREQU. DEP.} & {\it ENERGY DEP.} &
{\it AUTHOR} \\
 {\it SOURCE} & & {\it SCALE} & microwave & X-ray & \\
\hline
\hline
  {\bf Galaxy} & + & large & $\beta_{synch,ff,dust}$ & thermal & \\
 geometrical ? & ($-$) X-abs. & & & 0.3 keV & \\
\hline
 {\bf SZ thermal} & $-$ & ($<$ 10' clust.) & y - dist. & thermal & \\
 ({\bf clusters} / super- )& & ($< 5^\circ$ c-corr) & & 10 keV & \\
\hline
 {\bf SZ thermal} & $-$ & large & y - dist. & thermal & \cite{sutetal96} \\
 local group {\bf halo} & & & & 1 keV & \\
\hline
 X-ray/radio & + & small & flat, $\, \alpha < 0.5$ & $\gamma \approx 2
- 2.5 $ & \cite{fraetal89} \\
 point {\bf sources} & & & $(10 - 100 \, GHz)$ & $I = I_0
(E/E_0)^{-\gamma}$ & \\
\hline
 {\bf ISW / RS} &  $+$ & large & Planck & $\gamma \approx 2 - 2.5$ &
\cite{critur96}, \cite{kam96} \\
 in $\Lambda$ / open universe & & $\ell \approx 10$ & &  $I = I_0
(E/E_0)^{-\gamma}$ & \\
\hline
\end{tabular}
\caption{Overview of different effects introducing possible correlations
between microwave and X-ray data}
\end{table}
\vspace{1cm}
\normalsize
At the present state of observations, and with the available data, 
no astrophysically interesting correlation could be found. Still 
interesting limits can be drawn from this analysis, 
although no direct quantification of the effects. 
The current X-ray observations have much higher angular resolution and better 
signal to noise ratio than CMB measurements. But as these data were 
taken in the soft X-rays ($<$ 2 keV) they show predominantly a galactic signal
limiting a comparison with the XRB to small patches of the sky. 
This situation is going to change dramatically in the future. 
With the European CMB mission COBRAS/SAMBA and the German X-ray
satellite ABRIXAS, two all sky data sets will be available that have wide 
angular and spectral coverage and high angular and spectral
resolution. COBRAS/SAMBA \cite{tauthese},\cite{bouthese} will have 
9 frequency channels from 31 to 857 GHz with an angular resolution 
of $\sim$ 10 arcmin and a best sensitivity of $10^{-6}$ per
resolution element. 
ABRIXAS \cite{frietal96} will cover the energy range from 0.5 to 10 keV
with a sensitivity of 1.6 cts s$^{-1}$ for the XRB, a spectral resolution of
150 eV and an angular resolution of 1 arcmin. 
With these data sets a detailed statistical analysis for all the
effects discussed above will be feasible, including the direct comparison of
radio loud, X-ray luminous point sources and distant galaxy clusters.\\

\acknowledgements{I thank my collaborators from the ROSAT team,
R. Egger, G. Hasinger and J. Tr\"umper. The ROSAT project has
been supported by the Bundesministerium f\"ur Forschung und
Technologie (BMTF/DARA) and by the Max-Planck-Society. The COBE
datasets were developed by the NASA Goddard Space Flight Center 
under the guidance of the COBE Science Working Group and were 
provided by the NSSDC.}

\vfill
\end{document}